# The Distribution of Occupational Tasks in the United States: Implications for a Diverse and Aging Population

Samuel Cole,♣ Zachary Cowell,♦ John M. Nunley,♥ and R. Alan Seals♠

**Funding Details:** Research supported by Center for Financial Security Retirement and Disability Research Center funded by the Social Security Administration's Retirement and Disability Research Consortium

The research reported herein was performed pursuant to a grant from the U.S. Social Security Administration (SSA) funded as part of the Retirement and Disability Consortium. The opinions and conclusions expressed are solely those of the author(s) and do not represent the opinions or policy of SSA or any agency of the Federal Government. Neither the United States Government nor any agency thereof, nor any of their employees, makes any warranty, express or implied, or assumes any legal liability or responsibility for the accuracy, completeness, or usefulness of the contents of this report. Reference herein to any specific commercial product, process or service by trade name, trademark, manufacturer, or otherwise does not necessarily constitute or imply endorsement, recommendation or favoring by the United States Government or any agency thereof.

The authors thank seminar participants at the Center for Financial Security (University of Wisconsin-Madison) and at the Southern Economic Association. Any errors are our own.

♣ Cole, Graduate Student, Department of Economics, Auburn University, Auburn, AL 36849-5049, email: sdc0038@auburn.edu.

♦ Cowell, Graduate Student, Department of Economics, Auburn University, Auburn, AL 36849-5049, email: zlc0004@auburn.edu.

♥ Nunley, Ph.D., Chair and Professor, Department of Economics, University of Wisconsin—La Crosse, La Crosse, WI 54601, email: jnunley@uwlax.edu.

♠ **Corresponding Author: Seals, Ph.D.,** Associate Professor, Department of Economics, Auburn University, Auburn, AL 36849-5049, email: alan.seals@auburn.edu.

# The Distribution of Occupational Tasks in the United States: Implications for a Diverse and Aging Population

**Abstract**

We document the age-race-gender intersectionality in the distribution of occupational tasks in the United States. We also investigate how the task content of work changed from the early-2000s to the late-2010s for different age-race/ethnicity-gender groups. Using the Occupation Information Network (O*NET) and pooled cross-sectional data from the American Community Survey (ACS) we examine how the tasks that workers perform vary with age and over time. We find that White men transition to occupations high in non-routine cognitive tasks early in their careers, whereas Hispanic and Black men work mostly in physically demanding jobs over their entire working lives. Routine manual tasks increased dramatically for 55-67 year-old workers, except for Asian men and women. Policymakers will soon be challenged by financial stress on entitlement programs, reforms could have disproportionate effects on gender and racial/ethnic groups due to inequality in the distribution of occupational tasks.

# 1. Introduction

With further saturation of computer technology into daily life and increased exposure to international trade, work has changed dramatically in the 21st century (Acemoglu & Restrepo, 2020, 2021; D. Autor et al., 2021). In the United States and other developed countries, these extensive secular changes in the labor market coincide with a growing older population. In fact, one out of every four U.S. workers will be over the age of 55 by 2026 (Collins & Casey, 2017).[1]

Economic research on aging is primarily concerned with the health and financial well-being of older people. However, less is known about how the task content of work is distributed by age and across demographic groups.[2] Our study fills this void by (a) examining how the distribution of occupational task intensity in the U.S. varies with age for different racial/ethnic and gender group and (b) documenting how the mental, physical and social requirements of occupations have changed over time for different age-race/ethnicity-gender groups. Our analysis is based on data from the Occupation Information Network (O*NET) and the American Community Surveys (ACS) from 2005—2019. The longitudinal data on occupations from O*NET are linked to the ACS waves via a crosswalk between the Standard Occupation Classification (SOC) system and the O*NET occupation codes (referred to as O*NET-SOC codes). Once the data sets are combined, we produce employment-weighted statistics for the task-content measures from Acemoglu & Autor (2011), as in Autor et al. (2003) and Deming (2017).

---

[1] The age demographics may also be one of the drivers of automation. (Acemoglu & Restrepo, 2022) show that country's with older populations show an increase in automation, due to a shortage of middle-aged workers.

[2] Lahey & Oxley (2021) point out that the discrimination literature has focused on the intersectionality between race and gender, but not age. Lahey & Oxley (2021) find significant evidence of intersectional discrimination by age and race. In an eye-tracking laboratory experiment with randomized resumes, they find evidence that preferences for White-named job applicants, over Black-named applicants, fade as the job candidates are (artificially) aged.

The age profiles of different worker tasks reveal several insights, but perhaps the main takeaway is that men and women of different races/ethnicities perform starkly different types of work as they age. We find that White men transition to jobs with high non-routine cognitive analytical and non-routine cognitive interpersonal tasks around mid-career and they remain in jobs requiring high levels of these tasks until they reach retirement age. White women, by contrast, tend to perform more routine cognitive tasks, and the intensity at which routine cognitive tasks are undertaken rises consistently as they age. Among men, each racial/ethnic groups begin working in occupations with high levels of physical task intensity early in their careers. Although White men and Asian men transition out of physical work steadily as they age, the same pattern does not hold for Hispanic and Black men. In fact, these minority groups, especially Hispanic men, tend to work in the most physically demanding occupations from the start of their careers and this continues until reaching retirement age. For the most part, the same pattern holds among women. The only exception is the intensity at which physical tasks are completed starts out at a lower level than it does for men.

We next investigate how the distribution of tasks across age and other demographic groups has changed over the last decade and a half. Our analysis of the changes in worker tasks follows the approach of Ross (2017) and Atalay et al. (2020), as these studies examine both flows of workers across occupations and changes in tasks within occupations. We find that White men of all ages experienced increases in non-routine cognitive task intensity and that Black and Hispanic women of all ages experienced large increases in routine cognitive tasks. Non-routine physical tasks increased for 55–67 year-old men; however, Black and Hispanic men did not experience much change in physical tasks over the sample period and remain near the top of the distribution for all age groups. Most of the across-time variation in task intensity is the result of the occupations themselves changing in lieu of workers shifting from one occupation to another, a finding consistent with Atalay et al. (2021).

Policymakers concerned about impending financial pressures on entitlement programs, related to demographics and other structural forces in the economy, may find it useful to keep track of the kinds of tasks older people perform at work. A better understanding of how tasks are distributed across demographic groups will be useful to mitigate disproportionate effects on minorities associated with the coming policy changes to Social Security and other entitlement systems.

## 2. Data Sources

We follow Ross (2017, 2020) and create a panel of occupations from 2005—2019 using data from O*NET. We then rely on the taxonomy of tasks provided in Acemoglu and Autor (2011). In Table 1, we list the tasks, their abbreviations used later in the paper, the survey questions used to create the composite task measures, the O*NET module from which the survey questions were taken, and the measurement scale. When possible, we use the level scale, as it measures the degree to which a task/knowledge/skill/ability is required or needed to perform the tasks of a given occupation. We use the level scale from the abilities, skills, and work activities modules in lieu of the importance scale, as it provides a better measure of *how much* a task is completed for a given occupation. It is not always possible to use the level scale, as neither the level nor importance scales are available for the variables taken from the work context module. In this module, the units of measurement vary across survey questions. In general, the context scale is based on frequency and the time spent doing certain tasks. Following Autor, Levy, and Murnane (2003), we convert the average scores of the survey questions to a 0—10 scale, reflecting their weighted percentile rank in each year.

The next step is to link the panel of occupation characteristics to the American Community Survey (ACS). The ACS has been administered since 2000 on an annual basis, and O*NET survey began in 1998. We start our analysis is 2005 for two reasons. First, the ACS underwent several different sampling designs in the early years. In 2000, the sample was based on a 1-in-750 random draw from the US population. Sampling changed in 2001 to 2004 and varied between sampling of 1-in-232 and 1-in-261 random draw of the US population. In 2005 and subsequent years, the sampling changed to a 1-in-100 random sample of the US population. Second, online documentation from O*NET cautions against using the data in a longitudinal format prior to 2003.[3] Online that using the occupation information in a longitudinal format is problematic for prior years. We link the panel of occupations to the individual-level ACS data via a detailed occupation code and year. We limit the ACS sample to respondents between 16—67 years-old, who are employed and report a valid Standard Occupation Classification (SOC) code. The sample consists of 13,171,568 respondents. Using these data, we construct pseudo-lifecycle plots of worker tasks and measure the changes over time in task intensity for different age-race/ethnicity-gender groups.

When constructing the task measures, it is important to account for differences in the prevalence of occupations and, consequently, task content. For example, almost 50 percent of workers are employed in four major occupation groups: office and administration, sales, management, and healthcare practitioners. To account for differences in occupational prevalence, we follow Autor, Levy, and Murnane (2003) and construct labor-supply weights based on the number of hours usually worked and weeks worked during the previous year.[4] These weights are used in the analyses presented in the next two sections.

---

[3] As sensitivity checks, we checked our findings to the inclusion of 2003 and 2004 data from both O*NET and the ACSs. The inclusion of these additional data points does not affect our findings in a material way.

## 3. Task-Age Profiles

Using the individual-level ACS-O*NET linked data, we construct pseudo-lifecycle plots of the task intensities in which task intensities are plotted over age. The age profiles reflect the distribution of the task content of work in the U.S. labor market across the age distribution between 2005 and 2019. The task-age profiles are constructed for different workers from varying demographic backgrounds. In particular, we present task-age profiles by gender, race/ethnicity, and gender × race/ethnicity. The process to produce the task-age profiles involves three steps. The first step is to aggregate to the appropriate level. In our case, we conduct three separate aggregations to compute average task intensities by (a) age and gender (b) age and race/ethnicity, and (c) age, gender, and race/ethnicity. The sample sizes post-aggregation are 104, 364, and 520, respectively. The second step is to group the task intensity variables into centiles (100 bins), which creates a 0—100 scale. The third and final step is to fit a LOWESS curve through the data points with the task-intensity percentile measured on the $y$-axis and age measured on the $x$-axis.

From Figure 1, one observes that younger men and women are similar in terms of all task intensity variables, except those that require physical activity. For men, non-routine cognitive, whether analytical or interpersonal, tends to reach a plateau around mid-career, but they remain in occupations with high levels of intensity in these tasks until they reach retirement age. By contrast, these intensities rise for women until around mid-career before falling steadily as retirement age approaches. The intensity at which women complete routine cognitive tasks rises with age, but the opposite is true for men. Workers, particularly men, tend to work in occupations with more extensive physical demands when younger but tend to transition out of performing these tasks as they age.

---

[4] Because the measure for weeks worked in the ACS is a factor variable in lieu of a continuous measure, we use the mid-point in the ranges for each of the measure's values (1—6). The vast majority workers report working between 50 and 52 weeks per year. As such, the weight for these workers would equal hours usually worked times 51 (the midpoint).

In Figure 2, we analyze the task intensity measures by race/ethnicity and age. We focus on four racial/ethnic groups: White, Black, Hispanic, and Asian.[5] From the panels in Figure 2, it is apparent that different racial/ethnic groups perform different types of tasks over their working lives. White and Asian workers experience significant upticks in the extent to which they perform non-routine cognitive tasks, and these intensities rise until the mid-30s to mid-40s and then decline thereafter. Black and Hispanic workers tend to work in occupations that are low in non-routine cognitive analytical and non-routine cognitive interpersonal task intensity. Instead, they tend to work in occupations that are high in physical task intensity. In fact, Hispanic workers are employed in occupations that are near the top of the physical task intensity over their entire working lives. Physical task intensity tends to fall for Black workers with age, but the decline in physical task intensity with age is slower than it is for White and Asian workers. Lastly, for routine cognitive task intensity, we observe all races/ethnicities tending to work in more cognitively routine occupations at young ages. This pattern continues until mid-career and it stabilizes for Asian and Black workers but reverses for White and Hispanic workers and continues to decline as retirement age approaches.

---

[5] For our findings with respect to race/ethnicity, we are unable to follow the Office of Management and Budget (OMB) standards, as the sample sizes for the American Indian and Alaska Native (AIAN) and Native Hawaiian or Other Pacific Islanders (NHOPI) groupings are small and, therefore, potentially unreliable. We, however, follow their definitions to identify the four racial/ethnic groups used in our analysis.

In Figure 3, we present task-age profiles by race/ethnicity and gender. The solid lines represent men, and the dashed lines represent women. For non-routine cognitive analytical and non-routine cognitive interpersonal task intensity, the task-age proviles for all race/ethnicity-gender groups, except White men, reveal a concave relationship. For White men, these intensities rise and either plateau or continue to rise until reaching retirement age. The other race/ethnicity-gender groups experience rising intensity in these tasks, but that pattern tends to reverse around mid-career. For non-routine cognitive tasks, we also note the "shallowness" of the task-age profiles for Black and Hispanic men and women. Consider non-routine cognitive analytical task intensity. White men reach the 75$^{th}$ percentile around 40 years-old and then remain at a similar level of intensity as they age. Black and Hispanic workers, regardless of gender, reach approximately the 25$^{th}$ percentile around the mid-career, and then their task-age profiles plateau before falling as retirement age nears.

The intensity at which routine cognitive tasks are completed tends to rise in early career and then fall thereafter. The exceptions are White men and White women. For White men, routine cognitive task intensity begins around the 50$^{th}$ percentile but falls steadily until reaching the 1$^{st}$ percentile by age 67. White women, on the other hand, experience continually rising routine cognitive task intensity over their working lives: they begin around the 25$^{th}$ percentile, reach the 50$^{th}$ percentile around age 25, reach the 75$^{th}$ percentile around age 45, and then reach the 90$^{th}$ percentile in their early-60s.

In terms of physical work, we observe, for the most part, a negative relationship between the intensity of physical tasks, either routine manual or non-routine manual physical, and age. However, the rate at which the different racial/ethnic-gender groups transition out of physical work varies. White men and White women experience sharp drops in physical task intensity as they age.

Alternatively, the rate at which the task intensity declines for Black and Hispanic people, both men and women, is much slower. For example, Hispanic men and Black men tend to work in some of the most physically demanding occupations in the economy throughout their entire working lives. Physical task intensity falls for Asian men initially but the pattern stabilizes early in their early-30s, and they remain around the median for both routine manual and non-routine manual physical task intensity for the remainder of their working lives.

## 4. Changes in the Distribution of Tasks Over Time

We make comparisons of task intensities across demographic groups and time by examining two years of data: 2005 and 2019. We use the individual-level ACS-O*NET linked data, as in section 3. However, we limit the sample to workers employed in occupations observed in both 2005 and 2019. Ensuring the occupations are observed in both years allows for changes in the task intensities within occupations as well as changes resulting from workers switching from one occupation to another. In general, two methods have been proposed to study the evolution of task content in the U.S. economy. The approach by Autor et al. (2003), which is based on changes in aggregate task demand caused by changes in employment across occupations over time, and the approaches by Atalay et al. (2020) and Ross (2017), which allow one to measure the sum of between-occupation shifts in employment, as in Autor et al. (2003), and changes in task intensity within the occupations themselves over time.

Our analysis, like those used in Atalay et al. (2020) and Ross (2017), gives the total change in task intensity over time in lieu of only the change in task intensity due solely to changes in employment shares across occupations over time.[6] We compute the employment-weighted mean of the task intensities for each age-race/ethnicity-gender group in each of the two years. The resulting data set consists of 832 age-race/ethnicity-gender-year observations (=52 age groups × 4 racial/ethnic groups × 2 genders × 2 years). Following these calculations, we then rank and assign the task intensities to percentiles (0—100) for each year. Due to the difficulty of presenting the results for each of the 51 age groups, we instead compute the median task intensity percentile for each race/ethnicity-gender-year group for following age groups: (i) 16-24, (ii) 25-34, (iii) 35-44, (iv) 45-54, and (v) 55-67.[7]

The median percentiles for each demographic group are reported in Table 2. In general, the discrepancies between the 2005 and 2019 percentiles vary widely across age, race/ethnicity, and gender. However, the largest discrepancies exist across racial/ethnic-gender groups. Rather than comment on each group, consider the 55-67 age group, White men tend to shift out of routine cognitive tasks into physical tasks (routine manual and non-routine physical). Hispanic men tend to shift out of non-routine cognitive tasks, both analytical and interpersonal, into routine cognitive, routine manual, and non-routine manual physical tasks. For Black men, the intensity at which cognitive and physical tasks are performed declines between the two years. Black and Hispanic women moved upward 36 percentiles and 29 percentiles, respectively, in the routine cognitive task intensity distribution between 2005 and 2019. Interestingly, Hispanic women move up the distribution for each of the five tasks. The same is largely true for Black women, except non-routine physical task intensity falls by 12 percentiles. The task intensities tend to rise for White women, but the change is relatively smaller than that for Black and Hispanic women.[8]

---

[6] The main difference in the two approaches is the use of a base year in Autor et al. (2003) and longitudinal data on occupations by Atalay et al. (2020) and Ross (2017). Autor et al.'s study is constrained by the lack of available data. Atalay et al. (2020) build on Autor et al.'s work by extracting the text from job advertisements posted in leading newspapers from 1950—2000.

[7] For example, consider non-routine cognitive analytical task intensity for 55-67 year-old White men. For 2005, the percentiles assigned to each age group range from 77—91. The median value for this group is 83 and the mode is 82. Thus, the percentile for non-routine cognitive analytical task intensity assigned to White men in the 55-67 year-old age group for 2005 is the median value (i.e. 83).

As a comparison to the overall changes from Table 2, we present the task intensities from 2005 and 2019 using the Autor et al. (2003) method, which captures changes in task intensity from workers changing occupations, in lieu of the occupations themselves changing. The results are presented in Table A1. In addition, we conduct a simple decomposition of the overall change into two components: the change from workers moving across occupations and the change from the occupations themselves evolving. These results are presented in Table A2. In lieu of commenting in detail on the supplementary calculations, we note that we find, as does Atalay et al. (2020), that changes in employment shares across occupations are an important part of the observed change. However, examining the within and between occupation changes in isolation can be misleading, as it is common for the within channel to dominate the between channel, and vice versa.

## 5. Conclusion

An extensive literature documents the relationship between occupational task intensity and labor market outcomes (e.g., D. Autor & Dorn, 2013; D. H. Autor et al., 2003; Deming, 2017). However, the distribution of labor market tasks by age and other demographics has received less attention (Hurst et al., 2021). With the exception of Hudomiet & Willis (2021), who focus on the computerization of occupations, to our knowledge, there has been no study focused specifically on the allocation of tasks for specific age-gender-race/ethnicity groups. Filling this void in our understanding of how work changes for people as they age will be particularly important for government policy.

---

[8] In Appendix Figures A1.a—A1.e, we plot the median percentiles for each demographic group but for the entire sample period in lieu of only two years (2005 and 2019). The five plots present the median task intensities for each race/ethnicity-gender group. The plots differ based on the age group under investigation.

We illustrate the age-specific distribution of occupational tasks in the United States from 2005-2019. Stark racial/ethnic and gender divides in the age-specific allocation of tasks offer a common theme from our results. As they age, White men work in the most cognitively demanding jobs, while White women work in the least physically demanding occupations. Black and Hispanic men maintain the highest physical task intensity, whereas White men appear to transition to more cognitively intensive work early in their careers and remain in jobs with similar cognitive intensity until late career. On average, Black men and Hispanic men perform jobs with similar physical task intensity throughout their careers.

We also show that the kinds of work people of all ages perform has changed dramatically during the first part of the 21st century. An increase in routine cognitive and routine manual task intensity, particularly for the oldest group of workers, poses a significant challenge for policy makers. Older workers, a group that already suffers from rampant labor market discrimination (Lahey, 2008; Lahey & Oxley, 2021; Neumark et al., 2015) and lower productivity because of health limitations, may be more likely to be displaced by automation than younger workers. The more physically intense occupations of Black and Hispanic men over age 55 may be particularly susceptible to technological change, such as robotics (e.g., see Acemoglu & Restrepo, 2021); however, lower physical task intensity may also lengthen their work lives.

Policymakers in the United States and other developed countries will soon be challenged by financial stress on entitlement programs, largely due to the exit of baby boomers from the labor force. The relationship between worker tasks and relevant employment outcomes, such as earnings and labor force participation, will play an important role in determining the effects of the coming SSA reforms. Gender and racial/ethnic inequality in the distribution of tasks could contribute to racial/ethnic inequality in later-life disability (Kelley-Moore & Ferraro, 2004) and retirement savings (e.g., see Tamborini & Kim, 2020). Because in the U.S., Black and Hispanic retirees rely more on Social Security payments, which also depend on the length of work-life and career earnings, adjustments to retirement age and benefits will likely have disproportionate effects on these groups.

**References**

Acemoglu, D., & Autor, D. (2011). Skills, tasks and technologies: Implications for employment and earnings. In *Handbook of labor economics* (Vol. 4, pp. 1043–1171). Elsevier.

Acemoglu, D., & Restrepo, P. (2020). Robots and jobs: Evidence from US labor markets. *Journal of Political Economy*, *128*(6), 2188–2244.

Acemoglu, D., & Restrepo, P. (2021). *Tasks, Automation, and the Rise in US Wage Inequality*. National Bureau of Economic Research.

Acemoglu, D., & Restrepo, P. (2022). Demographics and automation. *The Review of Economic Studies*, *89*(1), 1–44.

Atalay, E., Phongthiengtham, P., Sotelo, S., & Tannenbaum, D. (2020). The evolution of work in the United States. *American Economic Journal: Applied Economics*, *12*(2), 1–34.

Autor, D., & Dorn, D. (2013). The growth of low-skill service jobs and the polarization of the US labor market. *American Economic Review*, *103*(5), 1553–1597.

Autor, D., Dorn, D., & Hanson, G. H. (2021). *On the Persistence of the China Shock*. National Bureau of Economic Research.

Autor, D. H., Levy, F., & Murnane, R. J. (2003). The skill content of recent technological change: An empirical exploration. *The Quarterly Journal of Economics*, *118*(4), 1279–1333.

Collins, S. M., & Casey, R. P. (2017). America's aging workforce: Opportunities and challenges. *Report from the Special Committee on Aging United States Senate*.

Deming, D. J. (2017). The Growing Importance of Social Skills in the Labor Market*. *The Quarterly Journal of Economics*, *132*(4), 1593–1640. https://doi.org/10.1093/qje/qjx022

Hudomiet, P., & Willis, R. J. (2021). *Computerization, Obsolescence, and the Length of Working Life*. National Bureau of Economic Research.

Hurst, E., Rubinstein, Y., & Shimizu, K. (2021). *Task-Based Discrimination*. National Bureau of Economic Research.

Kelley-Moore, J. A., & Ferraro, K. F. (2004). The black/white disability gap: Persistent inequality in later life? *The Journals of Gerontology Series B: Psychological Sciences and Social Sciences*, *59*(1), S34–S43.

Lahey, J. N. (2008). Age, Women, and Hiring: An Experimental Study. *Journal of Human Resources*, *43*(1), 30–56.

Lahey, J. N., & Oxley, D. R. (2021). Discrimination at the Intersection of Age, Race, and Gender: Evidence from an Eye-Tracking Experiment. *Journal of Policy Analysis and Management*.

Neumark, D., Burn, I., & Button, P. (2015). *Is it harder for older workers to find jobs? New and improved evidence from a field experiment*. National Bureau of Economic Research.

Tamborini, C. R., & Kim, C. (2020). Are you saving for retirement? Racial/ethnic differentials in contributory retirement savings plans. *The Journals of Gerontology: Series B*, *75*(4), 837–848.

Table 1 – Task Intensity Measure Definitions

| Task | Abbreviation | Survey Questions | Module | Scale |
|---|---|---|---|---|
| Nonroutine Cognitive Analytical | NR COGA | Analyzing data/information | Work Activities | Level, 0—7 |
| | | Thinking creatively | Work Activities | Level, 0—7 |
| | | Interpreting information for others | Work Activities | Level, 0—7 |
| Nonroutine Cognitive Interpersonal | NR COGI | Establishing and maintaining personal relationships | Work Activities | Level, 0—7 |
| | | Guiding, directing, and motivating others | Work Context | Context, 0—7 |
| | | Coaching/developing others | Work Context | Context, 0—7 |
| Routine Cognitive | R COG | Importance of repeating same tasks | Work Context | Context, 0—5 |
| | | Importance of being exact or accurate | Work Context | Context, 0—5 |
| | | Structured vs. unstructured work (reverse scale) | Work Context | Context, 0—5 |
| Routine Manual | R MAN | Pace determined by speed of equipment | Work Activities | Level, 0—7 |
| | | Controlling machines and processes | Work Context | Context, 0—7 |
| | | Spend time making repetitive motions | Work Context | Context, 0—5 |
| Nonroutine Manual Physical | NR PHYS | Operating vehicles, mechanized devices, or equipment | Work Activities | Level, 0—7 |
| | | Spend time using hands to handle, control or feel objects, tools or controls | Work Context | Context, 0—5 |
| | | Manual dexterity | Abilities | Level, 0—7 |
| | | Spatial orientation | Abilities | Level, 0—7 |

*Notes*: The task measures are based on the taxonomy provided in Acemoglu and Autor (2011).

Table 2
Changes in Task Intensity Over Time by Age Group, Race/Ethnicity, and Gender

| | Men | | | | | | | | | | | | Women | | | | | | | | | | | |
|---|---|---|---|---|---|---|---|---|---|---|---|---|---|---|---|---|---|---|---|---|---|---|---|---|
| | White | | | Black | | | Asian | | | Hispanic | | | White | | | Black | | | Asian | | | Hispanic | | |
| | 2005 | 2019 | Δ | 2005 | 2019 | Δ | 2005 | 2019 | Δ | 2005 | 2019 | Δ | 2005 | 2019 | Δ | 2005 | 2019 | Δ | 2005 | 2019 | Δ | 2005 | 2019 | Δ |
| **16-24** | | | | | | | | | | | | | | | | | | | | | | | | |
| NR COGA | 17 | 14 | -3 | 12 | 7 | -5 | 13 | 13 | 0 | 14 | 11 | -3 | 11 | 8 | -3 | 13 | 9 | -4 | 14 | 9 | -5 | 14 | 8 | -6 |
| NR COGI | 14 | 12 | -2 | 8 | 7 | -1 | 8 | 16 | 8 | 14 | 8 | -6 | 16 | 12 | -4 | 11 | 12 | 1 | 17 | 11 | -6 | 12 | 11 | -1 |
| R COG | 43 | 41 | -2 | 56 | 63 | 7 | 58 | 83 | 25 | 51 | 49 | -2 | 52 | 62 | 10 | 62 | 82 | 20 | 78 | 73 | -5 | 66 | 84 | 18 |
| R MAN | 93 | 94 | 1 | 92 | 92 | 0 | 63 | 75 | 12 | 97 | 94 | -3 | 35 | 59 | 24 | 38 | 55 | 17 | 26 | 46 | 20 | 47 | 61 | 14 |
| NR PHYS | 90 | 89 | -1 | 85 | 81 | -4 | 60 | 60 | 0 | 96 | 91 | -5 | 46 | 51 | 5 | 46 | 46 | 0 | 28 | 39 | 11 | 43 | 49 | 6 |
| **25-34** | | | | | | | | | | | | | | | | | | | | | | | | |
| NR COGA | 87 | 77 | -10 | 43 | 31 | -13 | 98 | 98 | 1 | 34 | 39 | 5 | 87 | 76 | -11 | 61 | 38 | -24 | 98 | 97 | -1 | 30 | 43 | 13 |
| NR COGI | 80 | 75 | -5 | 35 | 27 | -8 | 79 | 83 | 4 | 36 | 30 | -6 | 82 | 81 | -1 | 44 | 45 | 2 | 89 | 92 | 3 | 28 | 47 | 20 |
| R COG | 33 | 16 | -17 | 59 | 52 | -8 | 58 | 39 | -20 | 42 | 30 | -12 | 77 | 50 | -28 | 85 | 85 | 0 | 94 | 64 | -30 | 67 | 77 | 10 |
| R MAN | 66 | 59 | -7 | 74 | 74 | 1 | 42 | 16 | -26 | 93 | 82 | -11 | 12 | 16 | 5 | 23 | 34 | 11 | 23 | 3 | -20 | 40 | 34 | -6 |
| NR PHYS | 68 | 68 | 1 | 74 | 73 | -1 | 45 | 30 | -16 | 93 | 84 | -9 | 12 | 20 | 8 | 15 | 33 | 19 | 7 | 3 | -5 | 26 | 30 | 4 |
| **35-44** | | | | | | | | | | | | | | | | | | | | | | | | |
| NR COGA | 91 | 89 | -2 | 51 | 52 | 1 | 95 | 99 | 4 | 40 | 48 | 8 | 84 | 83 | -2 | 59 | 58 | -1 | 95 | 96 | 1 | 22 | 36 | 14 |
| NR COGI | 93 | 92 | -2 | 47 | 38 | -9 | 88 | 96 | 8 | 49 | 44 | -5 | 84 | 87 | 4 | 50 | 65 | 15 | 74 | 91 | 18 | 23 | 39 | 17 |
| R COG | 25 | 9 | -16 | 53 | 41 | -12 | 39 | 23 | -16 | 31 | 16 | -15 | 82 | 54 | -29 | 83 | 86 | 3 | 92 | 56 | -36 | 50 | 66 | 17 |
| R MAN | 63 | 48 | -15 | 77 | 71 | -6 | 45 | 11 | -35 | 90 | 87 | -3 | 18 | 7 | -11 | 27 | 24 | -3 | 37 | 4 | -34 | 53 | 44 | -9 |
| NR PHYS | 65 | 63 | -2 | 78 | 73 | -5 | 53 | 35 | -18 | 89 | 90 | 1 | 14 | 12 | -2 | 26 | 21 | -5 | 23 | 4 | -19 | 41 | 37 | -4 |
| **45-54** | | | | | | | | | | | | | | | | | | | | | | | | |
| NR COGA | 87 | 88 | 1 | 47 | 47 | -1 | 74 | 93 | 19 | 43 | 46 | 3 | 84 | 77 | -8 | 54 | 59 | 6 | 76 | 80 | 4 | 21 | 26 | 5 |
| NR COGI | 91 | 92 | 1 | 44 | 44 | 0 | 73 | 84 | 12 | 46 | 43 | -4 | 86 | 76 | -10 | 50 | 61 | 12 | 65 | 75 | 10 | 22 | 31 | 9 |
| R COG | 21 | 8 | -13 | 47 | 33 | -14 | 38 | 28 | -10 | 33 | 16 | -17 | 88 | 68 | -20 | 74 | 88 | 14 | 87 | 69 | -18 | 45 | 58 | 13 |
| R MAN | 61 | 51 | -11 | 80 | 73 | -8 | 58 | 36 | -22 | 88 | 88 | -1 | 18 | 10 | -8 | 31 | 25 | -6 | 47 | 22 | -25 | 50 | 58 | 9 |
| NR PHYS | 63 | 64 | 1 | 82 | 77 | -5 | 58 | 51 | -7 | 87 | 92 | 5 | 16 | 9 | -7 | 30 | 19 | -11 | 39 | 18 | -21 | 39 | 43 | 4 |
| **55-67** | | | | | | | | | | | | | | | | | | | | | | | | |
| NR COGA | 83 | 84 | 1 | 43 | 34 | -9 | 66 | 68 | 2 | 34 | 33 | -1 | 65 | 70 | 5 | 31 | 39 | 8 | 68 | 57 | -11 | 14 | 19 | 5 |
| NR COGI | 89 | 86 | -3 | 38 | 27 | -11 | 59 | 57 | -2 | 40 | 32 | -8 | 67 | 69 | 2 | 35 | 42 | 7 | 60 | 56 | -4 | 14 | 21 | 7 |
| R COG | 13 | 5 | -8 | 46 | 39 | -7 | 31 | 35 | 4 | 16 | 21 | 5 | 81 | 80 | -1 | 56 | 85 | 29 | 76 | 79 | 3 | 21 | 57 | 36 |
| R MAN | 39 | 53 | 14 | 76 | 76 | 0 | 56 | 52 | -4 | 83 | 84 | 1 | 13 | 15 | 2 | 28 | 31 | 3 | 45 | 30 | -15 | 46 | 52 | 6 |
| NR PHYS | 54 | 67 | 13 | 82 | 78 | -4 | 55 | 58 | 3 | 82 | 88 | 6 | 9 | 11 | 2 | 35 | 23 | -12 | 38 | 26 | -12 | 35 | 40 | 5 |

*Notes*: The table presents the median task intensity percentile for each race/ethnicity-gender group from five age ranges (16-24, 25-34, 35-44, 45-54, and 55-67) in 2005 and 2019 as well as the change between the two years. Computing the median task intensity percentiles requires several steps. Using the individual-level ACS-O*NET linked data, we compute employment-weighted means of the task intensity measures for each age-race/ethnicity-gender group in each of the two years. The resulting data set consists of 832 age-race/ethnicity-gender-year observations (=52 age groups × 4 racial/ethnic groups × 2 genders × 2 years). Following these calculations, we then rank and assign the task intensities to percentiles (1—99) for each year. Within each of the five age groupings, we then compute the median task intensity percentiles for each race/ethnicity-gender group in each year.

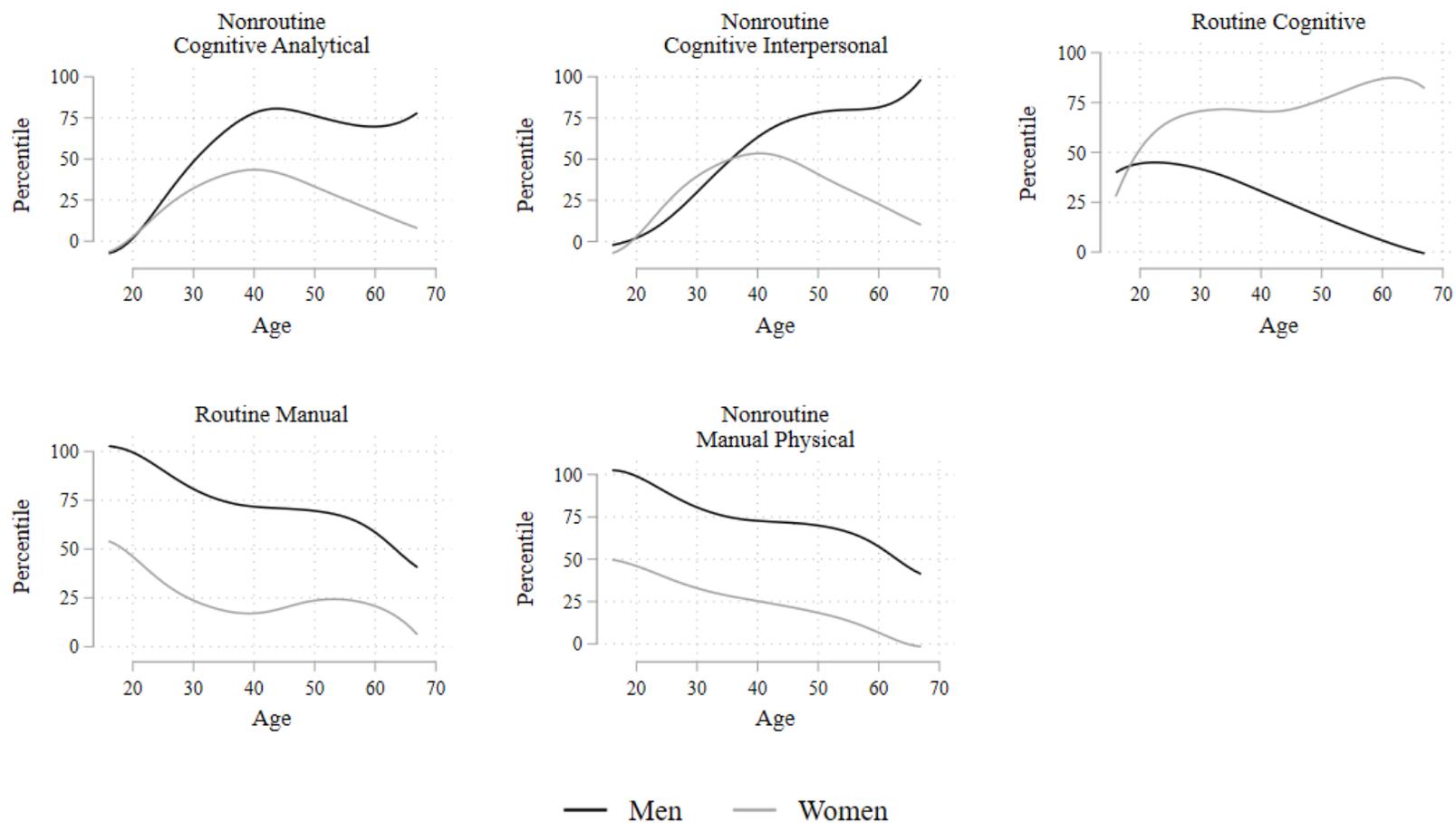

*Notes*: Using the combined O*NET and ACS data sets from 2004-2019 (described previously), the figure presents the relationship between the task intensity measures and age for men and women. To create the figure, we first aggregate the data to the 104 age-sex cells, and then separate the task intensity measures into centiles (100 bins). The plotted lines rely on a LOWESS fit with the bandwidth set at 0.8.

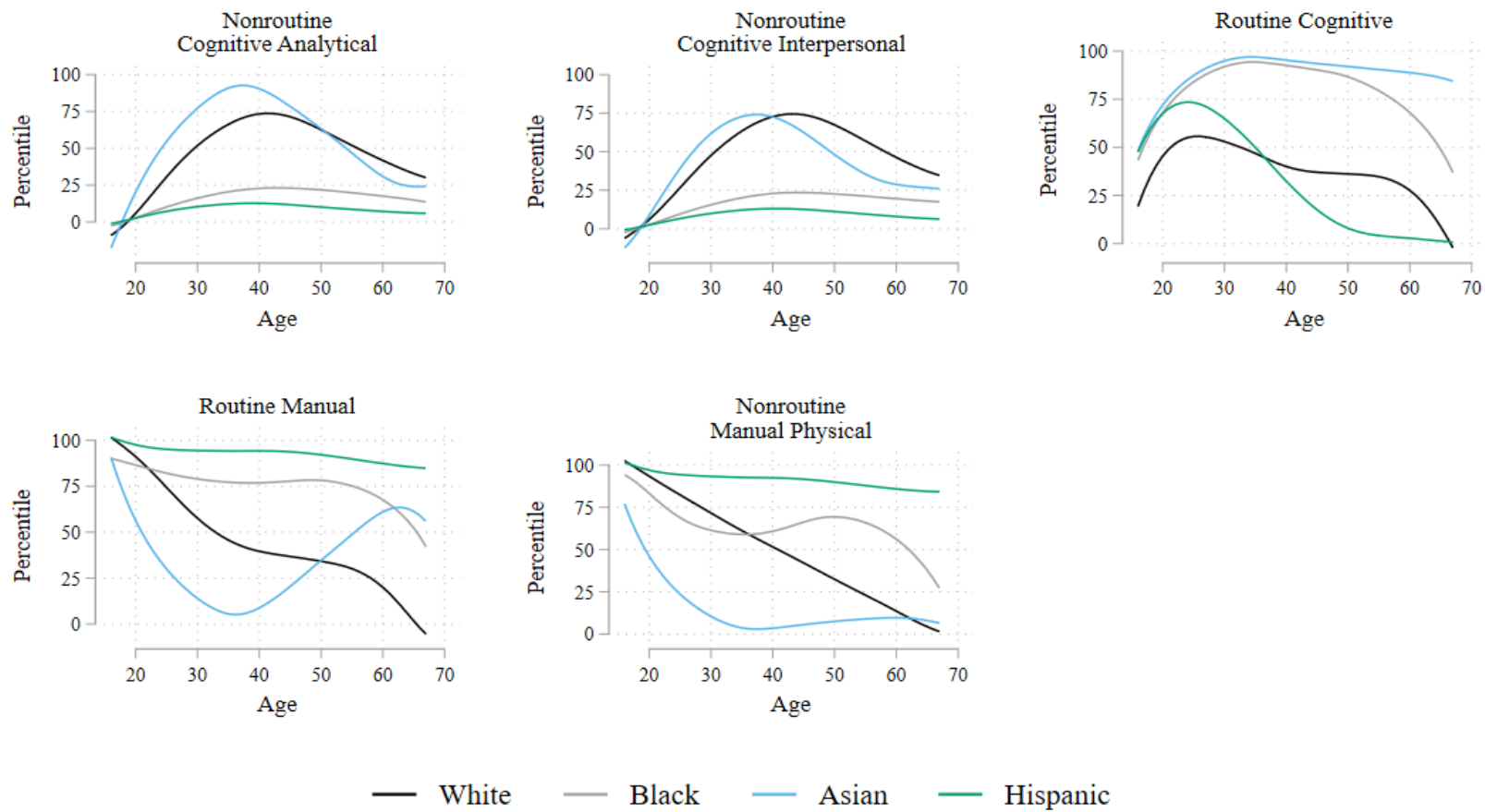

Notes: Using the combined O*NET and ACS data sets from 2004-2019 (described previously), the figure presents the relationship between the task intensity measures and age for different races/ethnicities. To create the figure, we first aggregate the data to the 364 age-race/ethnicity cells, and then separate the task intensity measures into centiles (100 bins). The plotted lines rely on a LOWESS fit with the bandwidth set at 0.8.

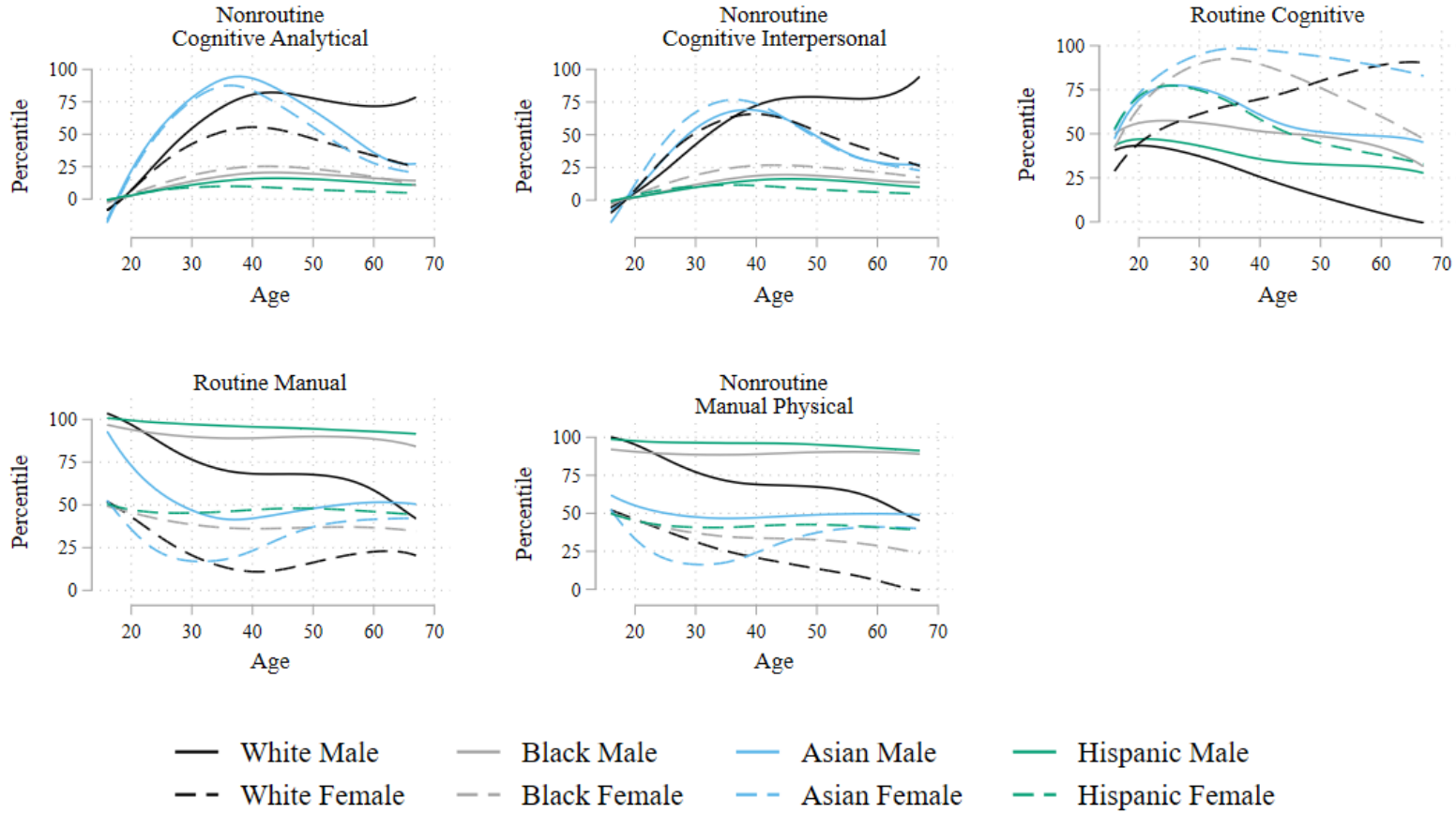

Figure 3
Task Intensities and Aging
by Gender and Race/Ethnicity

*Notes*: Using the combined O*NET and ACS data sets from 2004-2019 (described previously), the figure presents the relationship between the task intensity measures and age for different races/ethnicities. To create the figure, we first aggregate the data to the 520 age-race/ethnicity-gender cells, and then separate the task intensity measures into centiles (100 bins). The plotted lines rely on a LOWESS fit with the bandwidth set at 0.8.

# Appendix

Table A1
Between-Occupation Changes in Task Intensity Over Time by Age Group, Race/Ethnicity, and Gender

| | Men | | | | | | | | | | | | Women | | | | | | | | | | | |
|---|---|---|---|---|---|---|---|---|---|---|---|---|---|---|---|---|---|---|---|---|---|---|---|---|
| | White | | | Black | | | Asian | | | Hispanic | | | White | | | Black | | | Asian | | | Hispanic | | |
| | 2005 | 2019 | Δ | 2005 | 2019 | Δ | 2005 | 2019 | Δ | 2005 | 2019 | Δ | 2005 | 2019 | Δ | 2005 | 2019 | Δ | 2005 | 2019 | Δ | 2005 | 2019 | Δ |
| **16-24** | | | | | | | | | | | | | | | | | | | | | | | | |
| NR COGA | 17 | 18 | 1 | 12 | 8 | -4 | 13 | 14 | 1 | 14 | 12 | -2 | 11 | 9 | -2 | 13 | 9 | -4 | 14 | 9 | -5 | 14 | 8 | -6 |
| NR COGI | 14 | 15 | 1 | 8 | 6 | -2 | 8 | 14 | 6 | 14 | 9 | -5 | 16 | 12 | -4 | 11 | 9 | -2 | 17 | 11 | -6 | 12 | 9 | -3 |
| R COG | 43 | 54 | 11 | 56 | 55 | -1 | 58 | 70 | 12 | 51 | 56 | 5 | 52 | 13 | -39 | 62 | 23 | -39 | 78 | 43 | -35 | 66 | 33 | -33 |
| R MAN | 93 | 95 | 2 | 92 | 91 | -1 | 63 | 70 | 7 | 97 | 95 | -2 | 35 | 47 | 12 | 38 | 45 | 7 | 26 | 34 | 8 | 47 | 49 | 2 |
| NR PHYS | 90 | 94 | 4 | 85 | 86 | 1 | 60 | 69 | 9 | 96 | 95 | -1 | 46 | 51 | 5 | 46 | 47 | 1 | 28 | 44 | 16 | 43 | 49 | 6 |
| **25-34** | | | | | | | | | | | | | | | | | | | | | | | | |
| NR COGA | 87 | 80 | -8 | 43 | 34 | -10 | 98 | 97 | -1 | 34 | 41 | 7 | 87 | 79 | -8 | 61 | 39 | -23 | 98 | 99 | 1 | 30 | 38 | 8 |
| NR COGI | 80 | 75 | -5 | 35 | 33 | -2 | 79 | 82 | 3 | 36 | 41 | 5 | 82 | 84 | 3 | 44 | 42 | -2 | 89 | 96 | 7 | 28 | 40 | 13 |
| R COG | 33 | 38 | 5 | 59 | 71 | 12 | 58 | 72 | 14 | 42 | 50 | 8 | 77 | 59 | -18 | 85 | 78 | -7 | 94 | 93 | -1 | 67 | 75 | 8 |
| R MAN | 66 | 64 | -2 | 74 | 78 | 5 | 42 | 37 | -5 | 93 | 83 | -10 | 12 | 14 | 3 | 23 | 31 | 9 | 23 | 15 | -8 | 40 | 31 | -9 |
| NR PHYS | 68 | 67 | -1 | 74 | 76 | 3 | 45 | 40 | -6 | 93 | 85 | -8 | 12 | 20 | 8 | 15 | 35 | 20 | 7 | 6 | -1 | 26 | 29 | 3 |
| **35-44** | | | | | | | | | | | | | | | | | | | | | | | | |
| NR COGA | 91 | 91 | 0 | 51 | 57 | 6 | 95 | 99 | 4 | 40 | 50 | 10 | 84 | 85 | 1 | 59 | 59 | 0 | 95 | 96 | 1 | 22 | 27 | 5 |
| NR COGI | 93 | 92 | -2 | 47 | 48 | 1 | 88 | 99 | 11 | 49 | 57 | 8 | 84 | 90 | 7 | 50 | 60 | 10 | 74 | 94 | 20 | 23 | 30 | 8 |
| R COG | 25 | 23 | -2 | 53 | 62 | 9 | 39 | 45 | 7 | 31 | 30 | -1 | 82 | 67 | -16 | 83 | 81 | -3 | 92 | 84 | -8 | 50 | 54 | 4 |
| R MAN | 63 | 57 | -6 | 77 | 73 | -4 | 45 | 33 | -12 | 90 | 88 | -2 | 18 | 8 | -10 | 27 | 24 | -3 | 37 | 14 | -23 | 53 | 39 | -14 |
| NR PHYS | 65 | 60 | -5 | 78 | 74 | -4 | 53 | 39 | -14 | 89 | 89 | 0 | 14 | 12 | -3 | 26 | 22 | -4 | 23 | 7 | -16 | 41 | 31 | -10 |
| **45-54** | | | | | | | | | | | | | | | | | | | | | | | | |
| NR COGA | 87 | 89 | 2 | 47 | 49 | 2 | 74 | 93 | 19 | 43 | 46 | 3 | 84 | 79 | -5 | 54 | 58 | 4 | 76 | 80 | 4 | 21 | 20 | -2 |
| NR COGI | 91 | 92 | 1 | 44 | 49 | 6 | 73 | 88 | 15 | 46 | 54 | 8 | 86 | 79 | -7 | 50 | 54 | 5 | 65 | 75 | 10 | 22 | 22 | -1 |
| R COG | 21 | 16 | -5 | 47 | 48 | 1 | 38 | 42 | 4 | 33 | 23 | -10 | 88 | 84 | -4 | 74 | 78 | 4 | 87 | 84 | -3 | 45 | 36 | -10 |
| R MAN | 61 | 58 | -3 | 80 | 75 | -6 | 58 | 52 | -6 | 88 | 87 | -1 | 18 | 9 | -9 | 31 | 22 | -9 | 47 | 26 | -21 | 50 | 50 | 0 |
| NR PHYS | 63 | 63 | -1 | 82 | 79 | -3 | 58 | 50 | -8 | 87 | 90 | 4 | 16 | 9 | -8 | 30 | 22 | -8 | 39 | 23 | -16 | 39 | 37 | -2 |
| **55-67** | | | | | | | | | | | | | | | | | | | | | | | | |
| NR COGA | 83 | 85 | 2 | 43 | 42 | -1 | 66 | 68 | 2 | 34 | 36 | 2 | 65 | 70 | 5 | 31 | 38 | 7 | 68 | 53 | -15 | 14 | 15 | 1 |
| NR COGI | 89 | 82 | -7 | 38 | 34 | -4 | 59 | 64 | 5 | 40 | 36 | -4 | 67 | 71 | 4 | 35 | 36 | 1 | 60 | 48 | -12 | 14 | 16 | 2 |
| R COG | 13 | 10 | -3 | 46 | 48 | 2 | 31 | 35 | 4 | 16 | 23 | 7 | 81 | 91 | 10 | 56 | 72 | 16 | 76 | 74 | -2 | 21 | 25 | 4 |
| R MAN | 39 | 54 | 15 | 76 | 77 | 1 | 56 | 64 | 8 | 83 | 86 | 3 | 13 | 11 | -2 | 28 | 21 | -7 | 45 | 27 | -18 | 46 | 43 | -3 |
| NR PHYS | 54 | 60 | 6 | 82 | 84 | 2 | 55 | 63 | 8 | 82 | 88 | 6 | 9 | 9 | 0 | 35 | 23 | -12 | 38 | 24 | -14 | 35 | 33 | -2 |

*Notes*: The table presents the median task intensity percentile for each race/ethnicity-gender group from five age ranges (16-24, 25-34, 35-44, 45-54, and 55-67) in 2005 and 2019 as well as the change between the two years. Computing the median task intensity percentiles requires several steps. Using the individual-level ACS-O*NET linked data, we compute employment-weighted means of the task intensity measures for each age-race/ethnicity-gender group in each of the two years. However, the task intensities are not allowed to vary within occupations in these calculations. Instead, we use the values from a base year and assign those to all years. In our case, the base year is 2005. The resulting data set consists of 832 age-race/ethnicity-gender-year observations (=52 age groups × 4 racial/ethnic groups × 2 genders × 2 years). Following these calculations, we then rank and assign the task intensities to percentiles (1—99) for each year. Within each of the five age groupings, we then compute the median task intensity percentiles for each race/ethnicity-gender group in each year.

Table A2
Within and Between Changes in Task Intensity Over Time
by Age Group, Race/Ethnicity, and Gender

| | Men | | | | | | | | | | | | Women | | | | | | | | | | | |
|---|---|---|---|---|---|---|---|---|---|---|---|---|---|---|---|---|---|---|---|---|---|---|---|---|
| | White | | | Black | | | Asian | | | Hispanic | | | White | | | Black | | | Asian | | | Hispanic | | |
| | WI | BW | T | WI | BW | T | WI | BW | T | WI | BW | T | WI | BW | T | WI | BW | T | WI | BW | T | WI | BW | T |
| **16-24** | | | | | | | | | | | | | | | | | | | | | | | | |
| NR COGA | -4 | 1 | -3 | -1 | -4 | -5 | -1 | 1 | 0 | -1 | -2 | -3 | -1 | -2 | -3 | 0 | -4 | -4 | 0 | -5 | -5 | 0 | -6 | -6 |
| NR COGI | -3 | 1 | -2 | 1 | -2 | -1 | 2 | 6 | 8 | -1 | -5 | -6 | 0 | -4 | -4 | 3 | -2 | 1 | 0 | -6 | -6 | 2 | -3 | -1 |
| R COG | -13 | 11 | -2 | 8 | -1 | 7 | 13 | 12 | 25 | -7 | 5 | -2 | 49 | -39 | 10 | 59 | -39 | 20 | 30 | -35 | -5 | 51 | -33 | 18 |
| R MAN | -1 | 2 | 1 | 1 | -1 | 0 | 5 | 7 | 12 | -1 | -2 | -3 | 12 | 12 | 24 | 10 | 7 | 17 | 12 | 8 | 20 | 12 | 2 | 14 |
| NR PHYS | -5 | 4 | -1 | -5 | 1 | -4 | -9 | 9 | 0 | -4 | -1 | -5 | 0 | 5 | 5 | -1 | 1 | 0 | -5 | 16 | 11 | 0 | 6 | 6 |
| **25-34** | | | | | | | | | | | | | | | | | | | | | | | | |
| NR COGA | -3 | -8 | -10 | -3 | -10 | -13 | 1 | -1 | 1 | -2 | 7 | 5 | -3 | -8 | -11 | -1 | -23 | -24 | -2 | 1 | -1 | 5 | 8 | 13 |
| NR COGI | 1 | -5 | -5 | -6 | -2 | -8 | 1 | 3 | 4 | -11 | 5 | -6 | -4 | 3 | -1 | 3 | -2 | 2 | -4 | 7 | 3 | 7 | 13 | 20 |
| R COG | -22 | 5 | -17 | -20 | 12 | -8 | -34 | 14 | -20 | -20 | 8 | -12 | -10 | -18 | -28 | 7 | -7 | 0 | -29 | -1 | -30 | 3 | 8 | 10 |
| R MAN | -6 | -2 | -7 | -4 | 5 | 1 | -22 | -5 | -26 | -1 | -10 | -11 | 2 | 3 | 5 | 3 | 9 | 11 | -13 | -8 | -20 | 3 | -9 | -6 |
| NR PHYS | 2 | -1 | 1 | -4 | 3 | -1 | -10 | -6 | -16 | -1 | -8 | -9 | 1 | 8 | 8 | -2 | 20 | 19 | -4 | -1 | -5 | 1 | 3 | 4 |
| **35-44** | | | | | | | | | | | | | | | | | | | | | | | | |
| NR COGA | -2 | 0 | -2 | -5 | 6 | 1 | 0 | 4 | 4 | -3 | 10 | 8 | -2 | 1 | -2 | -1 | 0 | -1 | 0 | 1 | 1 | 9 | 5 | 14 |
| NR COGI | 0 | -2 | -2 | -10 | 1 | -9 | -3 | 11 | 8 | -13 | 8 | -5 | -3 | 7 | 4 | 5 | 10 | 15 | -3 | 20 | 18 | 9 | 8 | 17 |
| R COG | -14 | -2 | -16 | -21 | 9 | -12 | -22 | 7 | -16 | -14 | -1 | -15 | -13 | -16 | -29 | 5 | -3 | 3 | -28 | -8 | -36 | 13 | 4 | 17 |
| R MAN | -9 | -6 | -15 | -2 | -4 | -6 | -23 | -12 | -35 | -1 | -2 | -3 | -1 | -10 | -11 | 0 | -3 | -3 | -11 | -23 | -34 | 5 | -14 | -9 |
| NR PHYS | 3 | -5 | -2 | -1 | -4 | -5 | -4 | -14 | -18 | 1 | 0 | 1 | 1 | -3 | -2 | -2 | -4 | -5 | -3 | -16 | -19 | 6 | -10 | -4 |
| **45-54** | | | | | | | | | | | | | | | | | | | | | | | | |
| NR COGA | -1 | 2 | 1 | -3 | 2 | -1 | 0 | 19 | 19 | 0 | 3 | 3 | -3 | -5 | -8 | 2 | 4 | 6 | 0 | 4 | 4 | 7 | -2 | 5 |
| NR COGI | -1 | 1 | 1 | -6 | 6 | 0 | -4 | 15 | 12 | -12 | 8 | -4 | -3 | -7 | -10 | 7 | 5 | 12 | 0 | 10 | 10 | 9 | -1 | 9 |
| R COG | -8 | -5 | -13 | -15 | 1 | -14 | -14 | 4 | -10 | -8 | -10 | -17 | -16 | -4 | -20 | 10 | 4 | 14 | -15 | -3 | -18 | 23 | -10 | 13 |
| R MAN | -8 | -3 | -11 | -2 | -6 | -8 | -17 | -6 | -22 | 1 | -1 | -1 | 1 | -9 | -8 | 4 | -9 | -6 | -5 | -21 | -25 | 9 | 0 | 9 |
| NR PHYS | 2 | -1 | 1 | -2 | -3 | -5 | 1 | -8 | -7 | 2 | 4 | 5 | 1 | -8 | -7 | -3 | -8 | -11 | -5 | -16 | -21 | 6 | -2 | 4 |
| **55-67** | | | | | | | | | | | | | | | | | | | | | | | | |
| NR COGA | -1 | 2 | 1 | -8 | -1 | -9 | 0 | 2 | 2 | -3 | 2 | -1 | 0 | 5 | 5 | 1 | 7 | 8 | 4 | -15 | -11 | 4 | 1 | 5 |
| NR COGI | 4 | -7 | -3 | -7 | -4 | -11 | -7 | 5 | -2 | -4 | -4 | -8 | -2 | 4 | 2 | 6 | 1 | 7 | 8 | -12 | -4 | 5 | 2 | 7 |
| R COG | -5 | -3 | -8 | -9 | 2 | -7 | 0 | 4 | 4 | -2 | 7 | 5 | -11 | 10 | -1 | 13 | 16 | 29 | 5 | -2 | 3 | 32 | 4 | 36 |
| R MAN | -1 | 15 | 14 | -1 | 1 | 0 | -12 | 8 | -4 | -2 | 3 | 1 | 4 | -2 | 2 | 10 | -7 | 3 | 3 | -18 | -15 | 9 | -3 | 6 |
| NR PHYS | 7 | 6 | 13 | -6 | 2 | -4 | -5 | 8 | 3 | 0 | 6 | 6 | 2 | 0 | 2 | 0 | -12 | -12 | 2 | -14 | -12 | 7 | -2 | 5 |

*Notes*: This table decomposes the overall task intensity changes between 2005 and 2019 into "within" and "between" components. The former captures how the occupations themselves have changed over time, and the latter measures changes in task intensities due to workers moving from one occupation to another between the two years. The column headings "WI", "BW", and "T" indicate denote the within change, the between change, and the overall or total change. The statistics presented under the "BW" heading are the changes from Appendix Table A1, and those presented under the "T" heading are the changes from Table 2. The "WI" component is computed by subtracting the change in Appendix Table A1 from the change in Table 2. Netting out the between-occupation change leaves the within-occupation change, which is presented under the "WI" heading.

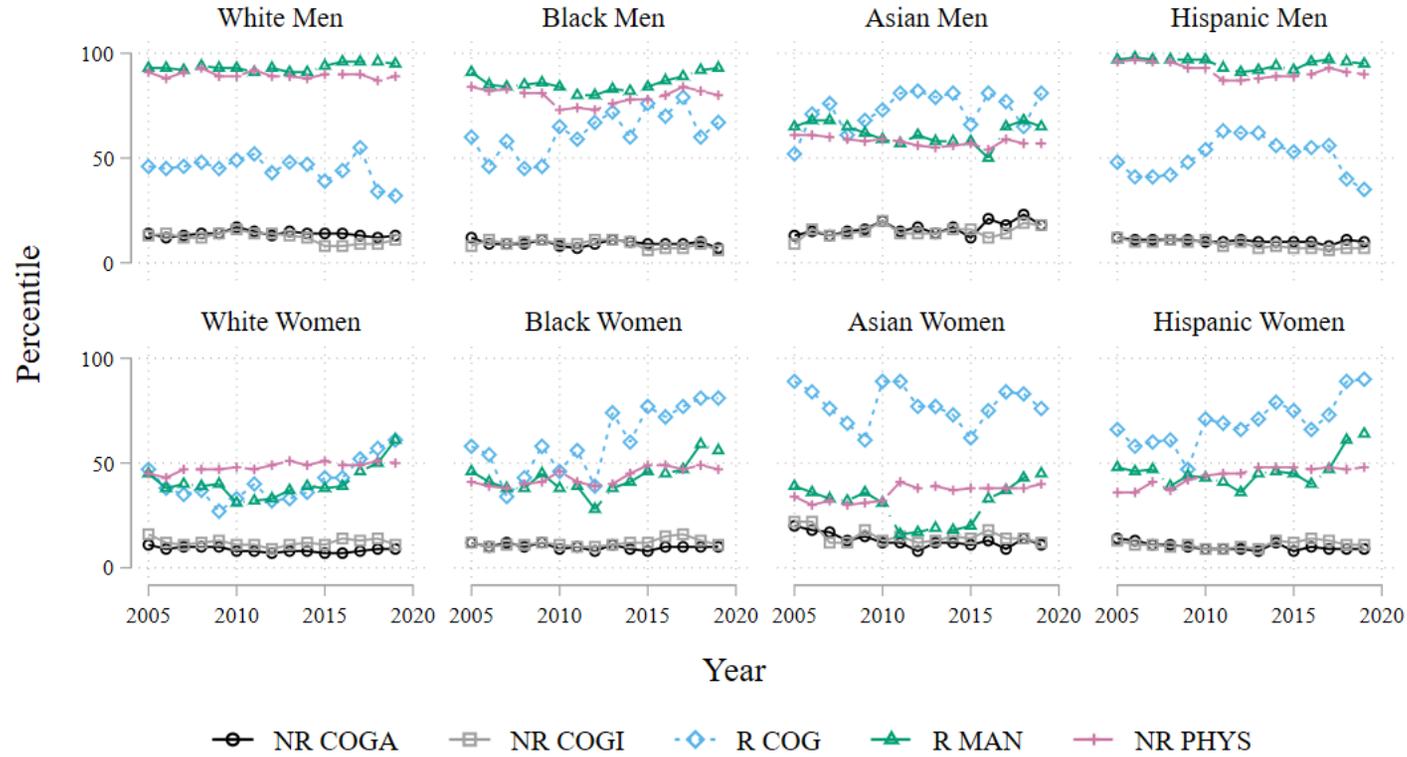

Figure A1.a

Changes in Task Intensity Within and Between Occupations Over Time by Race/Ethnicity, and Gender for 18-24 Year-Olds

*Notes*: The figure presents time plots from 2005-2019 of the median task intensity percentiles for each race/ethnicity-gender group for 16-24 year-olds. Computing the median task intensity percentiles requires several steps. Using the individual-level ACS-O*NET linked data, we compute employment-weighted means of the task intensity measures for each age-race/ethnicity-gender group in each of the two years. The resulting data set consists of 832 age-race/ethnicity-gender-year observations (=52 age groups × 4 racial/ethnic groups × 2 genders × 2 years). Following these calculations, we then rank and assign the task intensities to percentiles (1—99) for each year. With the 16-24 year-old age group, we then compute the median task intensity percentiles assigned to each race/ethnicity-gender group in each year.

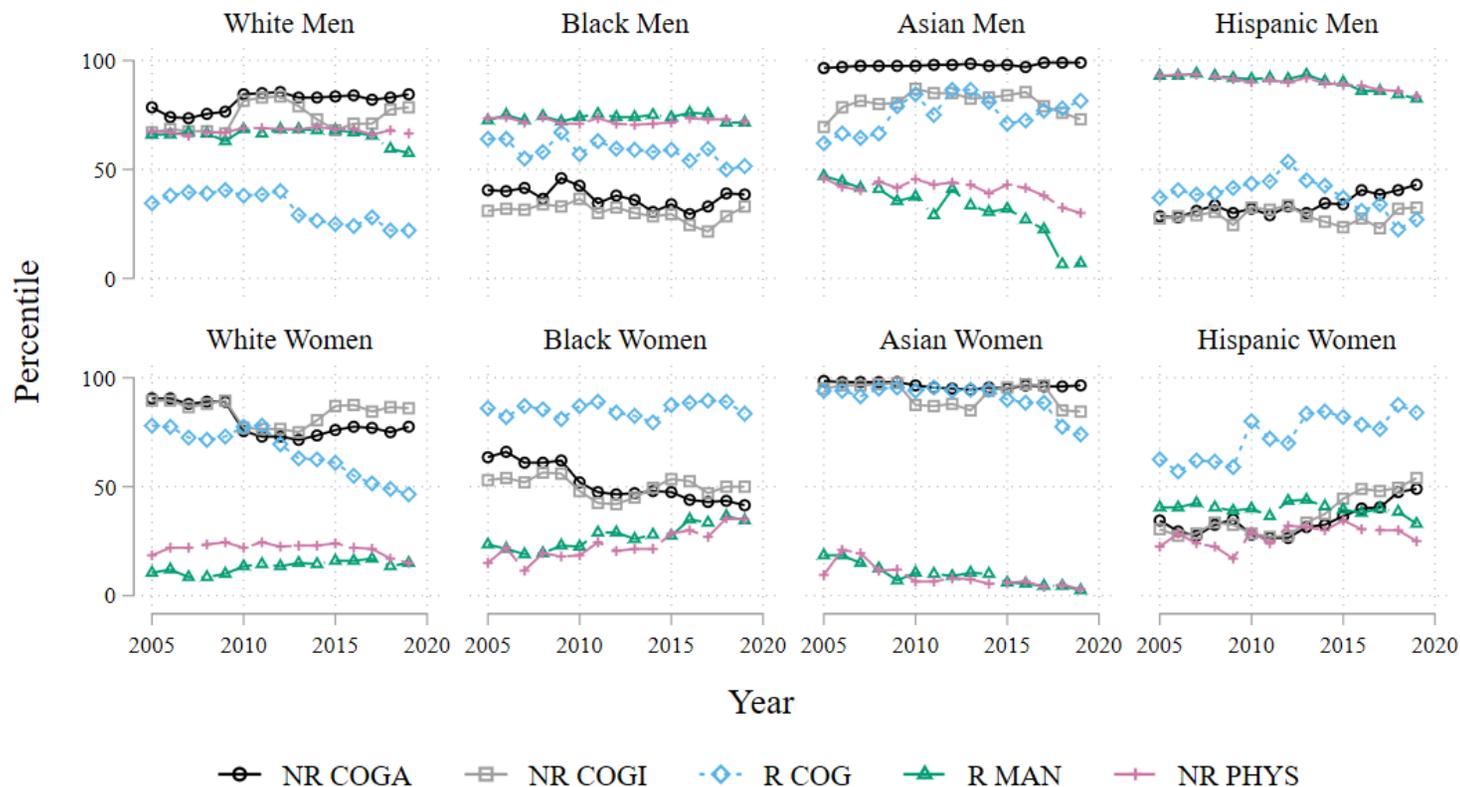

Figure A1.b
Changes in Task Intensity Within and Between Occupations Over Time by Race/Ethnicity, and Gender for 25-34 Year-Olds

*Notes*: The figure presents time plots from 2005-2019 of the median task intensity percentiles for each race/ethnicity-gender group for 25-34 year-olds. Computing the median task intensity percentiles requires several steps. Using the individual-level ACS-O*NET linked data, we compute employment-weighted means of the task intensity measures for each age-race/ethnicity-gender group in each of the two years. The resulting data set consists of 832 age-race/ethnicity-gender-year observations (=52 age groups × 4 racial/ethnic groups × 2 genders × 2 years). Following these calculations, we then rank and assign the task intensities to percentiles (1—99) for each year. Within the 25-34 year-old age group, we then compute the median task intensity percentiles assigned to each race/ethnicity-gender group in each year.

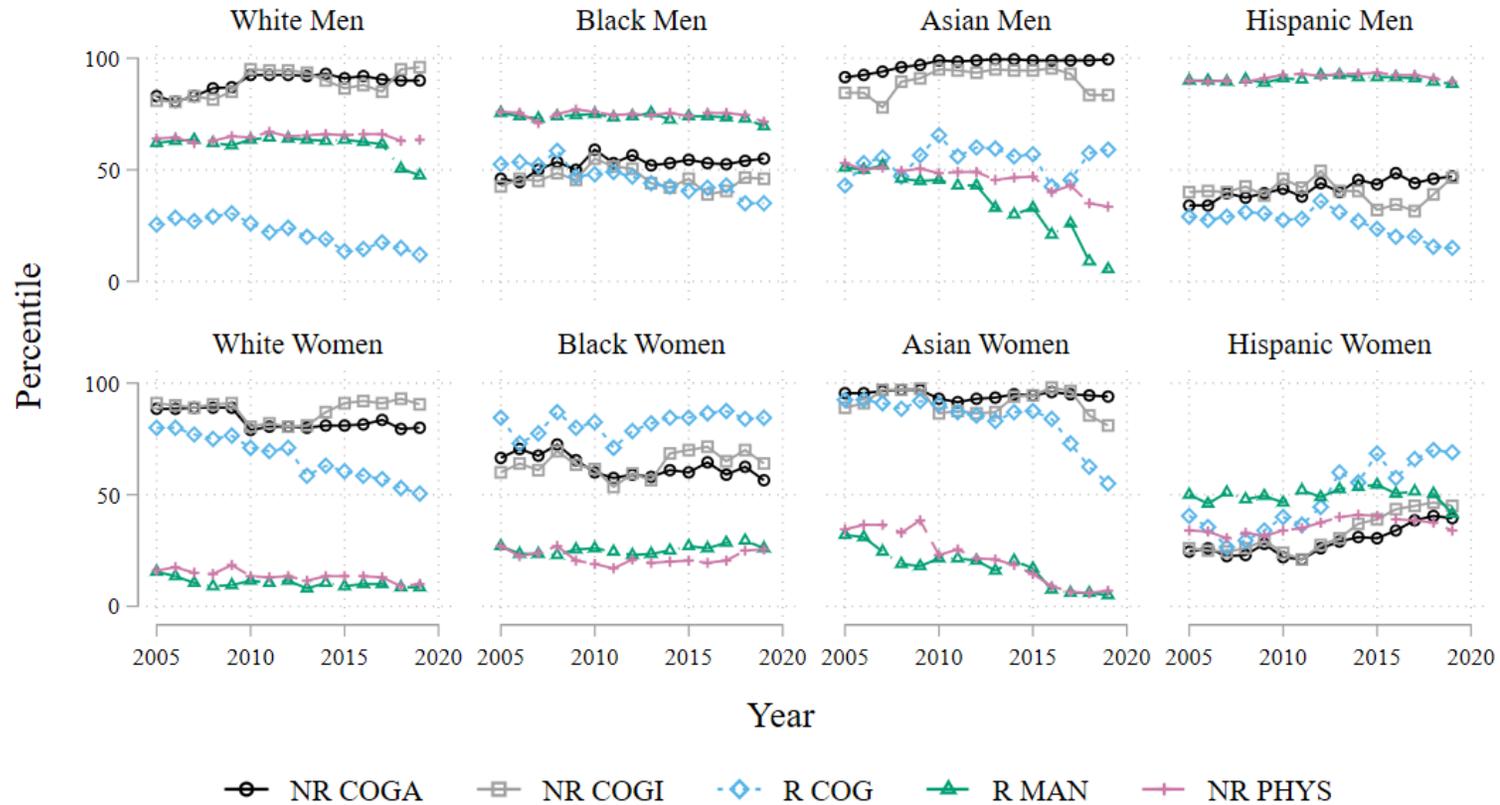

Figure A1.c
Changes in Task Intensity Within and Between Occupations Over Time
by Race/Ethnicity, and Gender for 35-44 Year-Olds

*Notes*: The figure presents time plots from 2005-2019 of the median task intensity percentiles for each race/ethnicity-gender group for 35-44 year-olds. Computing the median task intensity percentiles requires several steps. Using the individual-level ACS-O*NET linked data, we compute employment-weighted means of the task intensity measures for each age-race/ethnicity-gender group in each of the two years. The resulting data set consists of 832 age-race/ethnicity-gender-year observations (=52 age groups × 4 racial/ethnic groups × 2 genders × 2 years). Following these calculations, we then rank and assign the task intensities to percentiles (1—99) for each year. Within the 35-44 year-old age group, we then compute the median task intensity percentiles assigned to each race/ethnicity-gender group in each year.

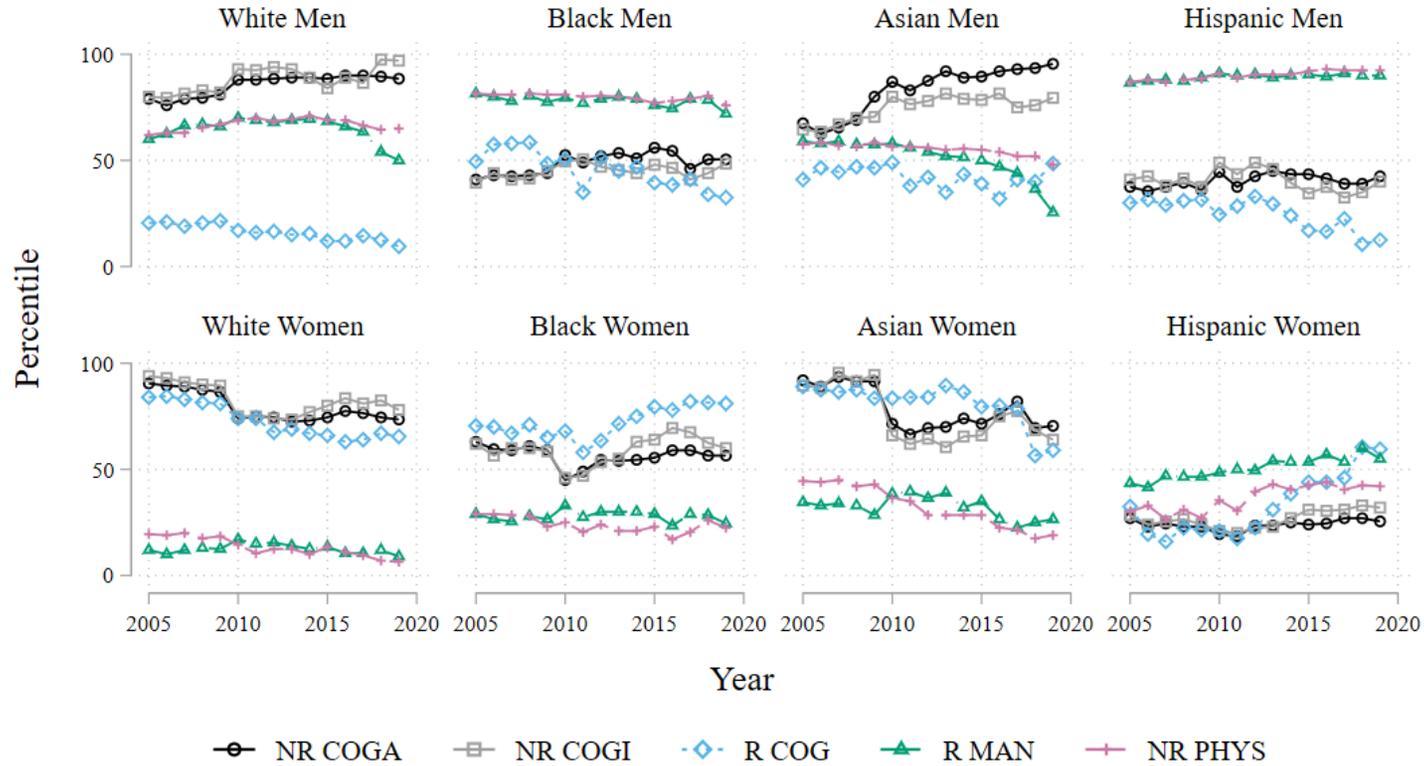

Figure A1.d

Changes in Task Intensity Within and Between Occupations Over Time by Race/Ethnicity, and Gender for 45-54 Year-Olds

*Notes*: The figure presents time plots from 2005-2019 of the median task intensity percentiles for each race/ethnicity-gender group for 45-54 year-olds. Computing the median task intensity percentiles requires several steps. Using the individual-level ACS-O*NET linked data, we compute employment-weighted means of the task intensity measures for each age-race/ethnicity-gender group in each of the two years. The resulting data set consists of 832 age-race/ethnicity-gender-year observations (=52 age groups × 4 racial/ethnic groups × 2 genders × 2 years). Following these calculations, we then rank and assign the task intensities to percentiles (1—99) for each year. Within the 45-54 year-old age group, we then compute the median task intensity percentiles assigned to each race/ethnicity-gender group in each year.

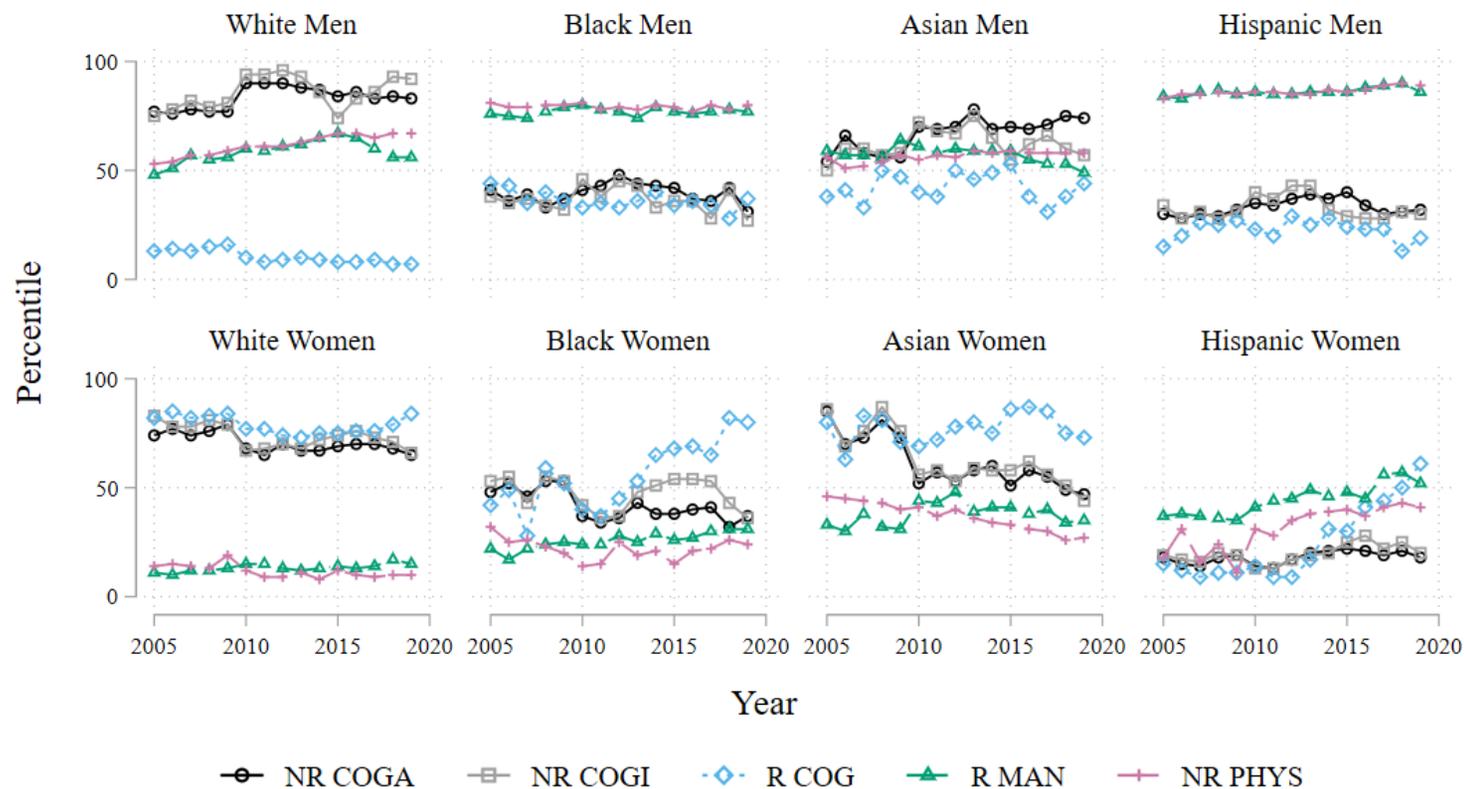

Figure A1.e
Changes in Task Intensity Within and Between Occupations Over Time by Race/Ethnicity, and Gender for 55-67 Year-Olds

*Notes*: The figure presents time plots from 2005-2019 of the median task intensity percentiles for each race/ethnicity-gender group for 55-67 year-olds. Computing the median task intensity percentiles requires several steps. Using the individual-level ACS-O*NET linked data, we compute employment-weighted means of the task intensity measures for each age-race/ethnicity-gender group in each of the two years. The resulting data set consists of 832 age-race/ethnicity-gender-year observations (=52 age groups × 4 racial/ethnic groups × 2 genders × 2 years). Following these calculations, we then rank and assign the task intensities to percentiles (1—99) for each year. Within the 55-67 year-old age group, we then compute the median task intensity percentiles assigned to each race/ethnicity-gender group in each year.